\documentclass[aps,prc,preprint,superscriptaddress,groupedaddress,
showpacs,floatfix]{revtex4}
\usepackage{graphicx}

\begin{document}

\title{Superscaling in Nuclei: A Search for Scaling Function Beyond the
Relativistic Fermi Gas Model}

\author{A.N. Antonov}
\affiliation{Institute of Nuclear Research and Nuclear Energy,
Bulgarian Academy of Sciences, Sofia 1784, Bulgaria}
\affiliation{Departamento de Fisica Atomica, Molecular y Nuclear,\\
Facultad de Ciencias  Fisicas, Universidad Complutense de
Madrid,\\
Madrid E-28040, Spain}

\author{M.K. Gaidarov}
\affiliation{Institute of Nuclear Research and Nuclear Energy,
Bulgarian Academy of Sciences, Sofia 1784, Bulgaria}

\author{D.N. Kadrev}
\affiliation{Institute of Nuclear Research and Nuclear Energy,
Bulgarian Academy of Sciences, Sofia 1784, Bulgaria}

\author{M.V. Ivanov}
\affiliation{Institute of Nuclear Research and Nuclear Energy,
Bulgarian Academy of Sciences, Sofia 1784, Bulgaria}

\author{E. Moya de Guerra}
\affiliation{Instituto de Estructura de la Materia, CSIC, Serrano
123,\\
28006 Madrid, Spain}

\author{J.M. Udias}
\affiliation{Departamento de Fisica Atomica, Molecular y Nuclear,\\
Facultad de Ciencias  Fisicas, Universidad Complutense de
Madrid,\\
Madrid E-28040, Spain}

\begin{abstract}
We construct a scaling function $f(\psi^{\prime})$ for inclusive
electron scattering from nuclei within the Coherent Density
Fluctuation Model (CDFM). The latter is a natural extension to
finite nuclei of the Relativistic Fermi Gas (RFG) model within
which the scaling variable $\psi^{\prime}$ was introduced by
Donnelly and collaborators. The calculations show that the
high-momentum components of the nucleon momentum distribution in
the CDFM and their similarity for different nuclei lead to
quantitative description of the superscaling in nuclei. The
results are in good agreement with the experimental data for
different transfer momenta showing superscaling for negative
values of $\psi^{\prime}$, including those smaller than --1.
\end{abstract}


\maketitle

\section{Introduction}
The $y$-scaling in the inclusive scattering of high-energy
electrons from nuclei has been actively studied in the last two
decades (e.g. \cite{Day90,Sick80,Pace82,Ciofi83,Ciofi99} and
references therein) following the idea from \cite{West75}. It has
been shown both theoretically and experimentally that a properly
defined function (scaling function) depends only on a single
variable $y$, the latter itself being a function of the
transferred momentum ${\bf q}$ and energy $\omega $
($y=y(q,\omega)$). It has been realized that at momenta $|{\bf
q}|>500$ MeV/c and energies $\omega $ at or below the quasielastic
peak position a nucleon is ejected in a "quasifree" way almost
without the effects of the strong interaction. This scaling is
usually called scaling of the first kind. It was shown that the
scaling function is sensitive to the high-momentum components of
the spectral function and nucleon momentum distribution. Thus its
knowledge can provide important information about the dynamical
ground-state properties, as well as about the reaction mechanism.

\begin{figure}[th]
\includegraphics[width=10cm]{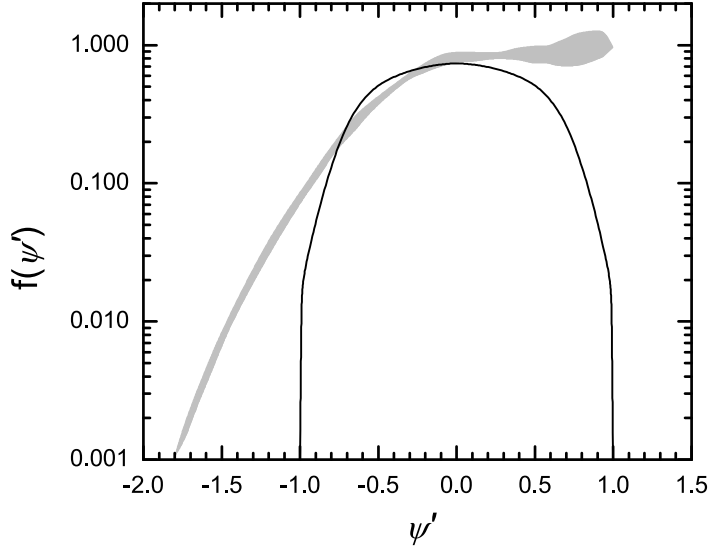}\\
\caption{Superscaling behaviour of inclusive electron-scattering.
The grey area represents experimental data \protect\cite{Donn99}
for $^{4}$He, $^{12}$C, $^{27}$Al and $^{197}$Au at $q=1000$
MeV/c. The solid line is the RFG scaling function calculated using
Eq. (\protect\ref{eq:24}) with $k_{F}=1.191$ fm$^{-1}$ from Ref.
\protect\cite{Donn99}.}
\label{fig1}
\end{figure}

The comparison of the scaling functions of various nuclei with
mass number $A\geq 4$ led to the conclusion that these functions
are the same \cite{Donn991,Donn99}. This behaviour is called
scaling of the second kind which, together with scaling of the
first kind, leads to superscaling. These studies followed the
theoretical concept of the superscaling introduced in
\cite{Barb98,Alber88} considering the properties of the
relativistic Fermi gas (RFG) model. The analyses of a large body
of inclusive scattering data for nuclei from $A$=4 to $A$=238 in
\cite{Donn991,Donn99} demonstrated that the data in the
low-$\omega $ side of the quasielastic peak exhibit superscaling
behaviour: the scaling functions are both independent on the
momentum transfer and on the mass number. In these analyses the
Fermi momentum for the RFG was used as a physical scale to define
the proper scaling variable $\psi^{\prime}$ for each nucleus. An
example of the superscaling behaviour of the inclusive
electron-scattering data for $q\approx 1000$ MeV/c and for the
$^{4}$He, $^{12}$C, $^{27}$Al, $^{56}$Fe and $^{197}$Au nuclei is
given in Fig. \ref{fig1} (the data are taken from Fig. 5 in
\cite{Donn99}). An important conclusion has been drawn in
\cite{Donn991,Donn99} that this universality is not restricted to
the region of the quasielastic peak ($|\psi^{\prime}|<1$) and that
the superscaling extends to larger values of $|\psi^{\prime}|$ and
thus, to the high-momentum components of the nucleon momentum
distributions in nuclei. The existence of high-momentum
components, and their similarity for different nuclei, is known to
be due to the short-range and tensor nucleon-nucleon correlations
(see, e.g. \cite{Ant88,Ant93} and references therein).

An extended study of scaling of the first and second kinds for
inclusive electron scattering from nuclei with emphasis on the
transverse response in the region above the quasielastic peak was
performed in \cite{Mai2002}. Approximate scaling of the second
kind was observed and its modest breaking was supposed to be due
to an inelastic version of the usual scaling variable. In
\cite{Barb2003} a unified relativistic approach used in the case
of quasielastic kinematics was applied to the analysis of
highly-inelastic electron scattering. The complete inelastic
spectrum was considered using the inelastic RFG model and its
phenomenological extension based on direct fits to data,
investigating the second-kind scaling behaviour as well.

As emphasized in \cite{Donn99}, the actual dynamical physics
content of the phenomenon considered is more complex than that
provided by the RFG framework. In particular, as noticed there,
the extension of the superscaling property to large negative
values of $\psi^{\prime}$ ($\psi^{\prime}<-1$) is not predicted by
the RFG model. This is seen in Fig. \ref{fig1} where we also give
a curve of the calculated RFG scaling function which is equal to
zero for $\psi^{\prime}\leq -1$. Thus, it is worth considering the
superscaling  in theoretical approaches which go beyond the RFG
model. One of them is the Coherent Density Fluctuation Model
(CDFM) (e.g. \cite{Ant79,Ant85,Ant88,Ant93}) which gives a natural
extension of the Fermi-gas case  to realistic finite nuclear
systems. The main aim of the present work is to see to what extent
superscaling can be explained using the CDFM. The theoretical
scheme is given in Section II, while the results and the
discussion are presented in Section III. The conclusions and the
final remarks are given in Section IV.

\section{The theoretical scheme}
\subsection{Basic relationships of the CDFM}
The CDFM was suggested \cite{Ant79} and developed
\cite{Ant85,Ant88,Ant93,Ant89} as a model for studying
characteristics of nuclear structure and nuclear reactions based
on the local density and momentum distributions as basic variables
of the theory and using the essential results of the infinite
nuclear matter theory. The model is related to the delta-function
limit of the generator coordinate method (GCM) \cite{Grif57}. In
the latter the total many-particle wave function $\Psi (\{{\bf
r}_{i}\})$ of a system of $A$ nucleons is written in a form of a
linear combination:
\begin{equation}
\Psi({\bf r}_{1},...,{\bf r}_{A})=\int F(x_{1},x_{2},...)\Phi({\bf
r}_{1},...,{\bf r}_{A};x_{1},x_{2},...)dx_{1}dx_{2}...,
\label{eq:1}
\end{equation}
where the generating function $\Phi(\{{\bf
r}_{i}\};x_{1},x_{2},...)$ depends on the radius-vectors of the
nucleons $\{{\bf r}_{i}\}$ (spin and isospin variables are
implied) and on the generator coordinates $x_{1},x_{2},...$. The
function $\Phi $ is usually chosen to be a Slater determinant
built up from single-particle wave functions corresponding to a
given construction potential parametrized by $x_{1},x_{2},...$.
The weight function $F(x_{1},x_{2},...)$ can be determined using
the variational principle as a solution of the Hill-Wheeler
integral equation:
\begin{equation}
\int [{\cal H}(x,x^{\prime})-E{\cal
I}(x,x^{\prime})]F(x^{\prime})dx^{\prime}=0.
\label{eq:2}
\end{equation}
In (\ref{eq:2}) the overlap kernel ${\cal I}(x,x^{\prime})$ and
the energy kernel ${\cal H}(x,x^{\prime})$ have the following
forms, respectively:
\begin{equation}
{\cal I}(x,x^{\prime})=\langle \Phi(\{{\bf r}_{i}\},x)|\Phi(\{{\bf
r}_{i}\},x^{\prime})\rangle,
\label{eq:3}
\end{equation}
\begin{equation}
{\cal H}(x,x^{\prime})=\langle \Phi(\{{\bf
r}_{i}\},x)|\hat{H}|\Phi(\{{\bf r}_{i}\},x^{\prime})\rangle,
\label{eq:4}
\end{equation}
where $i$=1,2,...,$A$, $\hat{H}$ is the Hamiltonian of the system
and $x$ denotes a set of $x_{1},x_{2},...$. For many-fermion
systems the kernels ${\cal I}(x,x^{\prime})$ and ${\cal
H}(x,x^{\prime})$ peak strongly at $x \sim x^{\prime}$
\cite{Wild77,Bauh80} and can be written in the form:
\begin{equation}
{\cal I}(x,x^{\prime})\simeq {\cal I}(x,x){\cal G}(x-x^{\prime}),
\label{eq:5}
\end{equation}
\begin{equation}
{\cal H}(x,x^{\prime})\simeq {\cal H}(x,x){\cal G}(x-x^{\prime}),
\label{eq:6}
\end{equation}
where ${\cal G}$ is peaked at $x \sim x^{\prime}$. It was shown in
\cite{Grif57} that the following delta-function approximation for
the kernels is valid in the GCM for the case of many-fermion
systems
\begin{equation}
{\cal I}(x,x^{\prime})\rightarrow \delta (x-x^{\prime}),
\label{eq:7}
\end{equation}
\begin{equation}
{\cal H}(x,x^{\prime})\rightarrow
-\frac{\hbar^{2}}{2m_{eff}}\delta^{\prime \prime}(x-x^{\prime})+V
\left(\frac{x+x^{\prime}}{2}\right )\delta (x-x^{\prime})
\label{eq:8}
\end{equation}
and that it leads to the Schr\"{o}dinger-type of equation
\cite{Grif57,Wild77,Dirac30} with an effective mass dependent on
the generator coordinate (see also \cite{Ant93}).

In the following we use for simplicity only one generator
coordinate. If the trial wave function $\Psi(\{{\bf r}_{i}\})$ in
(\ref{eq:1}) is normalized to the mass number $A$ and the weight
function is determined under the condition
\begin{equation}
\int_{0}^{\infty}|F(x)|^{2}dx=1,
\label{eq:9}
\end{equation}
then the delta-function approximation (\ref{eq:7}) leads to the
relationship:
\begin{equation}
\int \Phi^{*}(\{{\bf r}_{i}\},x^{\prime})\Phi(\{{\bf
r}_{i}\},x)d{\bf r}_{1}...d{\bf r}_{A}=A\delta(x-x^{\prime}).
\label{eq:10}
\end{equation}
Taking into account Eqs. (\ref{eq:5}) and (\ref{eq:10}) it was
suggested in the CDFM that a delta-function approximation leading
to that of Eq. (\ref{eq:10}) holds in the case of many-fermion
systems [14-16]:
\begin{equation}
\int \Phi^{*}({\bf r},{\bf r}_{2},...,{\bf r}_{A},x^{\prime})
\Phi({\bf r^{\prime}},{\bf r}_{2},...,{\bf r}_{A},x) d{\bf
r}_{2}...d{\bf r}_{A}\cong \rho_{x,x}({\bf r},{\bf
r^{\prime}})\delta(x-x^{\prime}).
\label{eq:11}
\end{equation}
In Eq. (\ref{eq:11}) $\rho_{x,x}({\bf r},{\bf r^{\prime}})$ is the
one-body density matrix corresponding to the wave function
$\Phi(\{{\bf r}_{i}\},x)$ which can be formally written as
\begin{eqnarray}
\rho_{x,x}({\bf r},{\bf r^{\prime}}) & \equiv & \rho_{x}({\bf
r},{\bf
r^{\prime}})\nonumber \\
&=& \frac{A}{\langle \Phi|\Phi \rangle} \int \Phi^{*}({\bf r},{\bf
r}_{2},...,{\bf r}_{A},x)\Phi({\bf r^{\prime}},{\bf
r}_{2},...,{\bf r}_{A},x)d{\bf r}_{2}...d{\bf r}_{A}.
\label{eq:12}
\end{eqnarray}
One can see that the integration of Eq. (\ref{eq:11}) (at ${\bf
r^{\prime}}$=${\bf r}$) over ${\bf r}$ using Eq. (\ref{eq:12})
leads to Eq. (\ref{eq:10}) which is the delta-function limit for
the overlap kernel in the GCM.

In the CDFM the generating function $\Phi(\{{\bf r}_{i}\},x)$
describes a system corresponding to a piece of nuclear matter with
a one-body density matrix (ODM) of the form
\begin{equation}
\rho_{x}({\bf r},{\bf r^{\prime}})=3\rho_{0}(x)
\frac{j_{1}(k_{F}(x)|{\bf r}-{\bf r^{\prime}}|)}{(k_{F}(x)|{\bf
r}-{\bf r^{\prime}}|)}\Theta \left (x-\frac{|{\bf r}+{\bf
r^{\prime}}|}{2}\right )
\label{eq:13}
\end{equation}
and uniform density
\begin{equation}
\rho_{x}({\bf r})=\rho_{0}(x)\Theta(x-|{\bf r}|),
\label{eq:14}
\end{equation}
where
\begin{equation}
\rho_{0}(x)=\frac{3A}{4\pi x^{3}}
\label{eq:15}
\end{equation}
and the generator coordinate $x$ is the radius of a sphere
containing all $A$ nucleons in it.

In Eq. (\ref{eq:13}) $j_{1}$ is the first-order spherical Bessel
function and
\begin{equation}
k_{F}(x)=\left(\frac{3\pi^{2}}{2}\rho_{0}(x)\right )^{1/3}\equiv
\frac{\alpha}{x} \;\;\;\; \mathrm{with} \;\;\;
\alpha=\left(\frac{9\pi A}{8}\right )^{1/3}\simeq 1.52A^{1/3}
\label{eq:16}
\end{equation}
is the Fermi momentum of such a piece of nuclear matter. Using
Eqs. (\ref{eq:1}) and (\ref{eq:11}) the ODM of the system in the
CDFM can be obtained as a superposition of the ODM's from Eq.
(\ref{eq:13}) \cite{Ant79,Ant85,Ant88,Ant93,Ant89}:
\begin{equation}
\rho({\bf r},{\bf r^{\prime}})=\int_{0}^{\infty}dx |F(x)|^{2}
\rho_{x}({\bf r},{\bf r^{\prime}}).
\label{eq:17}
\end{equation}
The Wigner distribution function which corresponds to the ODM from
Eq. (\ref{eq:17}) is:
\begin{equation}
W({\bf r},{\bf k})=\int_{0}^{\infty}dx|F(x)|^{2} W_{x}({\bf
r},{\bf k}),
\label{eq:18}
\end{equation}
where
\begin{equation}
W_{x}({\bf r},{\bf k})=\frac{4}{(2\pi)^{3}}\Theta (x-|{\bf
r}|)\Theta (k_{F}(x)-|{\bf k}|).
\label{eq:19}
\end{equation}
Then the density $\rho({\bf r})$ and the momentum distributions
$n({\bf k})$ in the CDFM are expressed by means of the same weight
function $|F(x)|^{2}$:
\begin{equation}
\rho({\bf r})=\int d{\bf k}W({\bf r},{\bf
k})=\int_{0}^{\infty}dx|F(x)|^{2}\frac{3A}{4\pi x^{3}}\Theta
(x-|{\bf r}|)
\label{eq:20}
\end{equation}
and
\begin{eqnarray}
n({\bf k})&=&\int d{\bf r}W({\bf r},{\bf k})=\frac{4}{(2\pi)^{3}}
\int_{0}^{\infty}dx|F(x)|^{2} \frac{4\pi x^{3}}{3}\Theta
(k_{F}(x)-|{\bf k}|)\nonumber \\
&=&\frac{4}{(2\pi)^{3}}\int_{0}^{\alpha/k}
dx|F(x)|^{2}\frac{4}{3}\pi x^{3},
\label{eq:21}
\end{eqnarray}
both normalized to the mass number:
\begin{equation}
\int \rho({\bf r})d{\bf r}=A; \;\;\;\; \int n({\bf k})d{\bf k}=A.
\label{eq:22}
\end{equation}
Different paths can be followed to find the function $F(x)$. Here,
instead of solving the differential equation from the
delta-function approximation to the Hill-Wheeler integral equation
(\ref{eq:2}), we adopt a convenient approach to the weight
function $F(x)$ proposed in \cite{Ant79,Ant85,Ant88,Ant93}. The
function $F(x)$ is obtained by means of a known density
distribution $\rho(r)$ for a given nucleus [from Eq.
(\ref{eq:20})]:
\begin{equation}
|F(x)|^{2}=-\frac{1}{\rho_{0}(x)} \left. \frac{d\rho(r)}{dr}\right
|_{r=x}, \;\;\;\;(\mathrm{at} \;\; d\rho(r)/dr\leq 0).
\label{eq:23}
\end{equation}

\subsection{The scaling function in the CDFM}
The scaling function in the RFG model expressed by the variable
$\psi^{\prime}$ has the form \cite{Donn99}
\begin{equation}
f_{RFG}(\psi^{\prime})=\frac{3}{4}(1-{\psi^{\prime}}^{2})\Theta
(1-{\psi^{\prime}}^{2})\frac{1}{\eta_{F}^{2}}\left \{
\eta_{F}^{2}+{\psi^{\prime}}^{2} \left[2+\eta_{F}^{2}-2\sqrt
{1+\eta_{F}^{2}}\right ]\right \},
\label{eq:24}
\end{equation}
where $\eta_{F}=k_{F}/m_{N}$, $m_{N}$ being the nucleon mass.

As shown in \cite{Donn99}, the relationship between
$\psi^{\prime}$ and the usual $y$-variable, in the approximation
for the mass of the residual nucleus $M_{A-1}^{0}\rightarrow
\infty$, is given by the expression
\begin{equation}
\psi^{\prime}=\frac{y}{k_{F}}\left
[1+\sqrt{1+\frac{1}{4\kappa^{2}}}\frac{1}{2}\eta_{F}\left
(\frac{y}{k_{F}}\right ) +{\cal O}[\eta_{F}^{2}]\right ],
\label{eq:25}
\end{equation}
where $\kappa=q/2m_{N}$.

Our basic assumption within the CDFM is that the scaling function
for a finite nucleus $f(\psi^{\prime})$ can be defined by means of
the weight function $|F(x)|^{2}$, weighting the scaling function
for the RFG at given $x$ (i.e. for a given density $\rho_{0}(x)$
(\ref{eq:15}) and Fermi momentum (\ref{eq:16})). Thus the scaling
function $f(\psi^{\prime})$ in the CDFM will be an infinite
superposition of the RFG scaling functions $f(\psi^{\prime}(x))$.

Let us introduce the notation
\begin{equation}
c\equiv \frac{1}{2m_{N}}\sqrt{1+\frac{1}{4\kappa^{2}}}.
\label{eq:26}
\end{equation}
Then one can write from Eqs. (\ref{eq:25}) and (\ref{eq:26}),
neglecting ${\cal O}[\eta_{F}^{2}]$, the scaling variable
$\psi^{\prime}_{x}(y)$ corresponding to the relativistic Fermi gas
with the density $\rho_{0}(x)$ (\ref{eq:15}) and Fermi momentum
$k_{F}(x)$ (\ref{eq:16}) in the form:
\begin{equation}
\psi^{\prime}_{x}(y)=\frac{p(y)}{k_{F}(x)}=\frac{p(y)x}{\alpha},
\label{eq:27}
\end{equation}
where for the cases of interest
\begin{equation}
p(y)=\left \{
\begin{array}{ll}
y(1+cy),      &\mbox{$\;\;\;\; y\geq 0$}\\
-|y|(1-c|y|), &\mbox{$\;\;\;\; y\leq 0$ ,$\;\;|y|\leq 1/(2c)$}.
\end{array}
\right.
\label{eq:28}
\end{equation}
For further use it is more convenient to introduce the notation
\begin{equation}
\psi^{\prime}_{x}(y)=\frac{k_{F}}{k_{F}(x)}\frac{p(y)}{k_{F}}=\frac{k_{F}}{k_{F}(x)}
\psi^{\prime}.
\label{eq:29}
\end{equation}
Using the $\Theta$-function in Eq. (\ref{eq:24}), the weighted
scaling function for a finite nucleus can be presented by the
integral
\begin{eqnarray}
f(\psi^{\prime})&=&\int_{0}^{\alpha/(k_{F}|\psi^{\prime}|)}dx
|F(x)|^{2}\frac{3}{4}\left[1-\left(
\frac{k_{F}x\psi^{\prime}}{\alpha}\right )^{2}\right ]\nonumber \\
& \times & \left \{1+\left (\frac{x m_{N}}{\alpha}\right )^{2}
\left( \frac{k_{F}x\psi^{\prime}}{\alpha}\right )^{2} \left
[2+\left (\frac{\alpha}{xm_{N}}\right )^{2}-2\sqrt{1+\left
(\frac{\alpha}{xm_{N}}\right )^{2}}\right ]\right \},
\label{eq:30}
\end{eqnarray}
where the Fermi momentum $k_{F}$ will not be a fitting parameter
(as it is in the RFG model) for the different nuclei, but will be
also calculated consistently in the CDFM,
\begin{equation}
k_{F}=\int_{0}^{\infty}dx
k_{F}(x)|F(x)|^{2}=\alpha\int_{0}^{\infty}dx \frac{1}{x}|{\cal
F}(x)|^{2}
\label{eq:31}
\end{equation}
for each nucleus, with $\alpha$ given by Eq. (\ref{eq:16}). As can
be seen from Eqs. (\ref{eq:30}), (\ref{eq:29}) and (\ref{eq:28})
in our approach the scaling function $f(\psi^{\prime})$ is
symmetric at the change of $\psi^{\prime}$ to -$\psi^{\prime}$ up
to $|\psi^{\prime}|=(4ck_{F})^{-1}$.

The scaling function $f(\psi^{\prime})$ has been calculated using
Eq. (\ref{eq:30}) by means of the weight function $|F(x)|^{2}$
determined from its relationship to the density distribution
$\rho(r)$ [Eq. (\ref{eq:23})]. For the latter we used those
obtained from experimental data on electron scattering from nuclei
and muonic atoms.

\section{Results of calculations and discussion}
We calculated the scaling function $f(\psi^{\prime})$
(\ref{eq:30}) for various nuclei and transfer momenta. A
symmetrized diffused Fermi density distribution has been used for
$^{4}$He and $^{12}$C \cite{Burov98} and a diffused Fermi
distribution for the heavier nuclei. The values of the half-radius
$R$ and diffuseness parameter $b$ are given in Table \ref{table1}
together with the results for the CDFM Fermi momentum $k_{F}$
(\ref{eq:31}).

\begin{table}
\centering \caption{Values of the parameters $R$ and $b$ (in fm)
used in the calculations and the results for $k_{F}$ (in
fm$^{-1}$) obtained in the CDFM.}
\begin{tabular}{cccccc}
\hline \hline
Nuclei & & $R$ & $b$ & $k_{F}$ \\
\hline
$^{4}$He                       & & 1.710 & 0.290 & 1.201 \\
$^{12}$C                       & & 2.470 & 0.420 & 1.200 \\
$^{27}$Al    \cite{Vries87}    & & 3.070 & 0.519 & 1.267 \\
$^{56}$Fe    \cite{Vries87}    & & 4.111 & 0.558 & 1.270 \\
$^{197}$Au   \cite{Patt2003}   & & 6.419 & 0.449 & 1.335 \\
\hline \hline
\end{tabular}
\label{table1}
\end{table}

\begin{figure}[th]
\includegraphics[width=12cm]{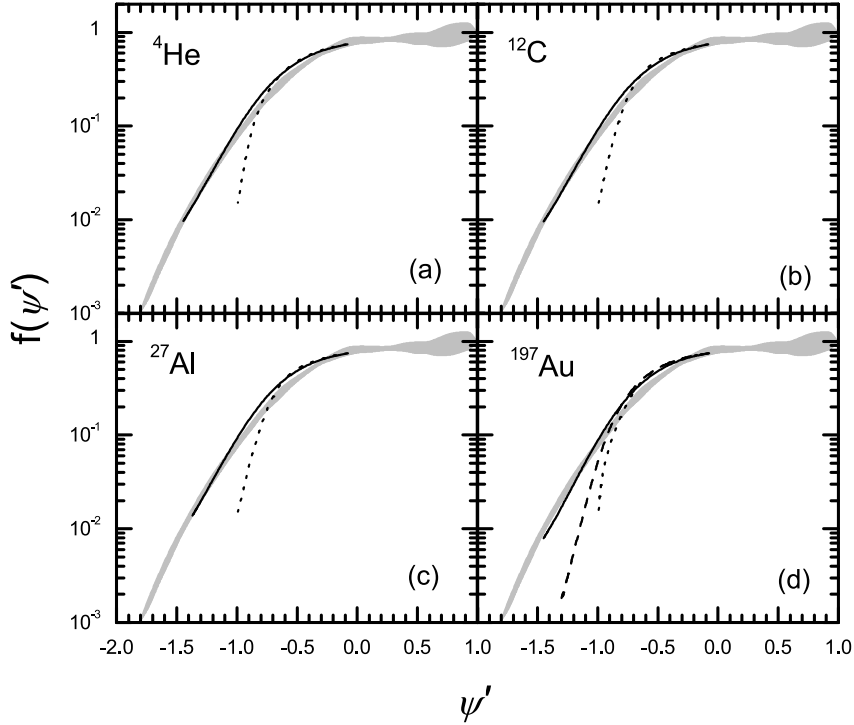}\\
\caption{Results for the scaling function in the CDFM (solid line)
calculated using Eqs. (\protect\ref{eq:30}) and
(\protect\ref{eq:31}) at $q=1000$ MeV/c and for $^{4}$He,
$^{12}$C, $^{27}$Al and $^{197}$Au (with $b=1.0$ fm for the
latter) compared with the data (grey area) from
\protect\cite{Donn99}. The dotted line is the RFG result using Eq.
(\protect\ref{eq:24}). The dashed line in the case of $^{197}$Au
corresponds to the CDFM result with $b=0.449$ fm.}
\label{fig2}
\end{figure}

The results for the scaling function are compared with the
experimental data from \cite{Donn991,Donn99} which are given in
our figures by a grey area. In Fig. \ref{fig2} are presented the
results for the scaling function in the CDFM for $q$=1000 MeV/c
and for $^{4}$He, $^{12}$C, $^{27}$Al and $^{197}$Au. The values
of the parameters $R$ and $b$ for $^{4}$He and $^{12}$C (given in
Table 1) lead to charge rms radii 1.71 fm and 2.47 fm,
respectively, which coincide with the experimental ones
\cite{Vries87}. The values of $R$ and $b$ for $^{27}$Al are taken
from \cite{Vries87}. The results of the CDFM scaling function
(solid lines) are compared with the RFG predictions (dotted
lines). In the RGF model, due to the $\Theta$-function in Eq.
(\ref{eq:24}), $f(\psi^{\prime})=0$ for $\psi^{\prime}\leq -1$. As
can be seen, the CDFM results give a good agreement with the data
for the interval for $\psi^{\prime}$ from 0 till
$\psi^{\prime}<-1$ for all nuclei (including $^{56}$Fe which is
not shown). The only exception was observed for $^{197}$Au using
the values of the parameters $R=6.419$ fm and $b=0.449$ fm given
in \cite{Patt2003}, for which the result is shown in Fig.
\ref{fig2} by dashed line.

At this point we would like to note that, generally, the weight
function $|F(x)|^{2}$ which is related to the density distribution
[Eqs. (\ref{eq:20}) and (\ref{eq:23})], is also related to the
momentum distribution [Eq. (\ref{eq:21})]. This connects through
Eq. (\ref{eq:30}) the scaling function $f(\psi^{\prime})$ with
$n(k)$. The deviation of $f(\psi^{\prime})$ from the data in the
case of $^{197}$Au ($b=0.449$) is due to the particular
$A$-dependence of $n(k)$ in the present approach [Eqs.
(\ref{eq:21}),(\ref{eq:23})].

\begin{figure}[th]
\includegraphics[width=9cm]{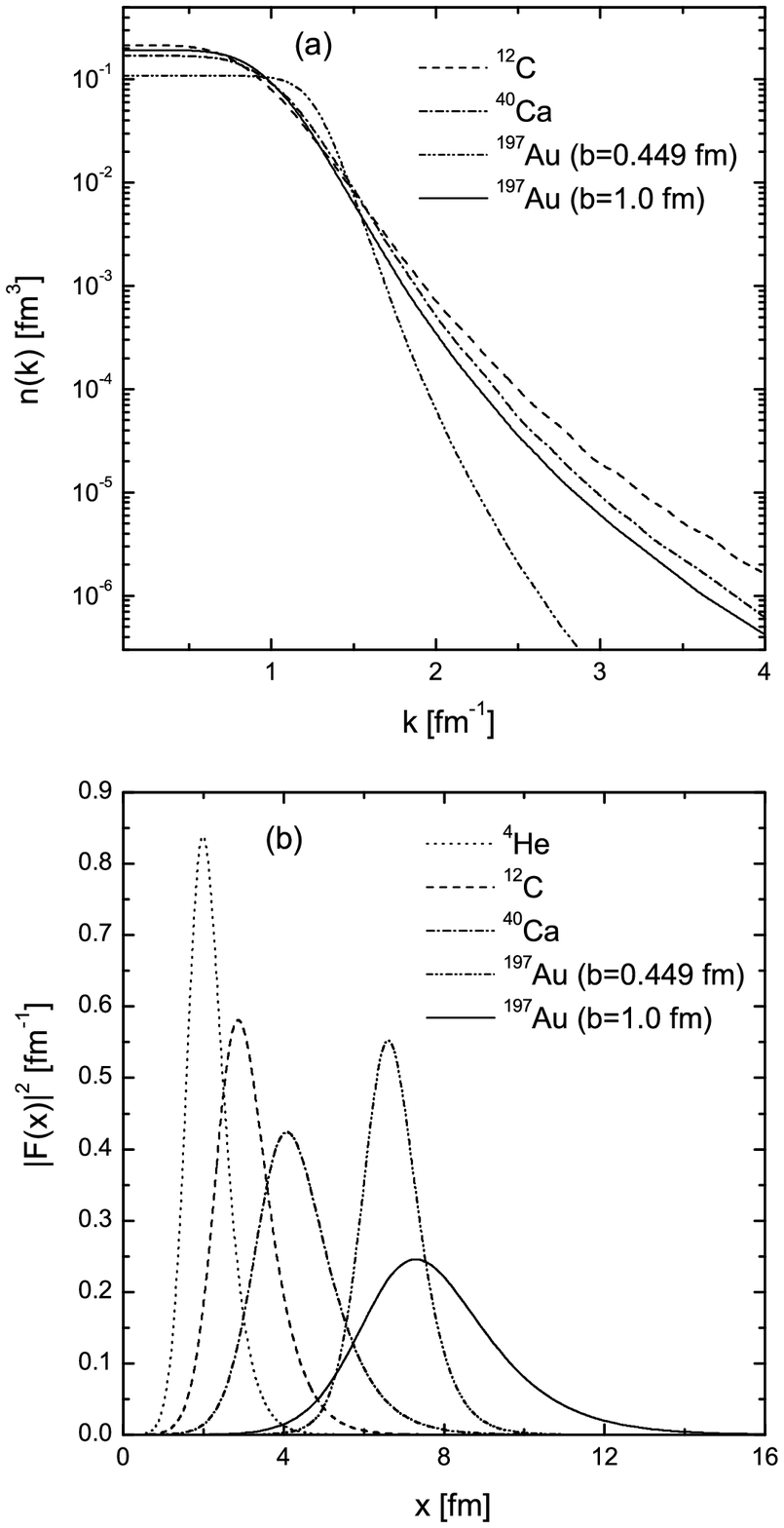}\\
\caption{(a) Nucleon momentum distribution $n(k)$ calculated in
the CDFM using Eqs. (\protect\ref{eq:21}) and
(\protect\ref{eq:23}) for $^{12}$C, $^{40}$Ca and $^{197}$Au (for
the latter with $b=0.449$ fm and $b=1.0$ fm); (b) The weight
function $|F(x)|^{2}$ of the CDFM calculated using Eq.
(\protect\ref{eq:23}) for $^{4}$He, $^{12}$C, $^{40}$Ca and
$^{197}$Au (for the latter with $b=0.449$ fm and $b=1.0$ fm).}
\label{fig3}
\end{figure}

To understand the origin of the agreement for light and medium
nuclei, and of the discrepancy for $^{197}$Au with $b=0.449$ fm,
we show in Fig. \ref{fig3} the momentum distributions $n(k)$
(\ref{eq:21}) (Fig. \ref{fig3}(a)) and the weight functions
$|F(x)|^{2}$ (Fig. \ref{fig3}(b)). As seen in Fig. \ref{fig3}(a),
for the $^{12}$C and $^{40}$Ca the CDFM momentum distributions
depend weakly on $A$ and have similar high-momentum ($k>1$
fm$^{-1}$) tails. Inspection of Eqs. (\ref{eq:21}) and
(\ref{eq:30}) (which involve the same weighting function
$|F(x)|^{2}$) shows that the interval of interest for
$\psi^{\prime}$ ($-2\leq \psi^{\prime}\leq 0$) corresponds to the
$k$ interval $0\leq k\leq -2.5$ fm$^{-1}$ in $n(k)$. Thus the
similar high-momentum tails of $n(k)$ in the light ($^{12}$C) and
medium ($^{40}$Ca) nuclei lead to the similar CDFM scaling curves
which are in agreement with the superscaling data.

The deviation between the CDFM scaling function and data in Fig.
\ref{fig2} for $^{197}$Au ($b=0.449$ fm) takes place at
$\psi^{\prime}\leq -1$ and corresponds to values of the momentum
$k\geq 3$ fm$^{-1}$. The CDFM high-momentum tail of $^{197}$Au for
$b=0.449$ fm is much smaller than that of light and medium nuclei
(see Fig. \ref{fig3}(a)) causing the above mentioned deviation. In
Fig. \ref{fig3}(b) we give also the weight function $|F(x)|^{2}$
calculated in the CDFM for $^{4}$He, $^{12}$C, $^{40}$Ca and
$^{197}$Au using Eq. (\ref{eq:23}) and the corresponding density
distributions mentioned above. As can be seen $|F(x)|^{2}$ is a
one-peak function which for $^{4}$He, $^{12}$C and $^{40}$Ca
follows a particular trend with increasing $A$. The strength of
the peak decreases (consequently its width increases) with
increasing $A$, and the peak is displaced to higher $x$ values.
This trend is broken by the behaviour of $|F(x)|^{2}$ for
$^{197}$Au when we take $b=0.449$ fm.

To improve the $A$-dependence of the momentum distribution in the
CDFM for the heaviest nucleus, we use a procedure that may be
somewhat artificial but which is useful to show the role of the
obtained, more realistic, new high-momentum components of $n(k)$
on the scaling function. This can be achieved by taking an
effective larger value of the parameter $b$ in the diffuse Fermi
density distribution which is used to obtain the weight function
$|F(x)|^{2}$ and, hence, the scaling function $f(\psi^{\prime})$.
We take the value $b=1.0$ fm, for which the high-momentum
components of $n(k)$ in $^{197}$Au are similar to those in light
and medium nuclei. This can be seen in Fig. \ref{fig3}(a) for
$^{197}$Au (solid line). The change of the behaviour of the weight
function $|F(x)|^{2}$ which leads to this shape of $n(k)$ for
$^{197}$Au can be seen in Fig. \ref{fig3}(b) (solid line). In this
case the function $|F(x)|^{2}$ follows the trend previously
observed for light and medium nuclei of decreasing strength
(increasing width) with increasing $A$. The values of $|F(x)|^{2}$
for $0\leq x \leq 5.2$ fm determine the behaviour of
$f(\psi^{\prime})$ for $-1.5<\psi^{\prime}<-1$. As can be seen
from Fig. \ref{fig3}(b) in this region $|F(x)|^{2}$ is quite
different for $b=0.449$ fm and for $b=1.0$ fm. The results of the
calculations of the scaling function for $^{197}$Au with $b=1.0$
fm are presented in Fig. \ref{fig2} by a solid line and, as can be
seen, they are in good agreement with the data. This confirms the
view that the behaviour of the scaling function is related to the
properties of the momentum distribution at large values of $k$
($k>1.5$ fm$^{-1}$), and that the similarity of the high-momentum
tails of $n(k)$ leads to the scaling of second kind.

Here we would like to note that the use of an effective value of
$b$ for $^{197}$Au can be merely seen as an artificial procedure
to improve the $A$-dependence of $n(k)$ for the heaviest nucleus.
This shows what would be the results of the CDFM for the scaling
function when the high-momentum tails of $n(k)$ are realistic,
even for the heaviest nucleus, and are similar to those of light
and medium nuclei. We do not imply that the actual diffuseness of
the density distribution of $^{197}$Au should be that large.
However, it is also worth pointing out that all the nucleons may
contribute to the scaling function for the transverse electron
scattering and that the diffuseness of the mass density for a
nucleus like $^{197}$Au may be larger than that of the charge
density.

\begin{figure}[th]
\includegraphics[width=10cm]{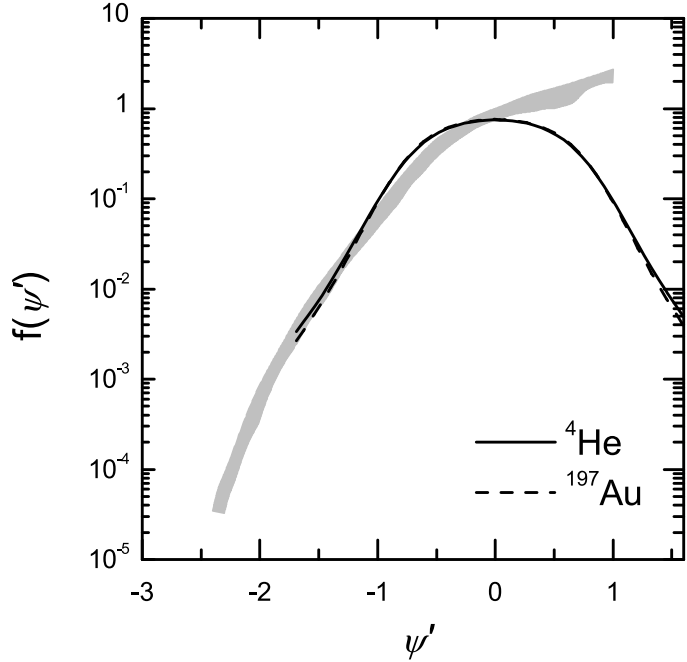}\\
\caption{Results of the CDFM for the superscaling functions of
$^{4}$He (solid line) and $^{197}$Au (dashed line) at $q=1650$
MeV/c compared with the experimental data (grey area) from
\cite{Donn99}.}
\label{fig4}
\end{figure}

The results for the scaling function in the CDFM in the case of
$q$=1650 MeV/c are given in Fig. \ref{fig4} for $^{4}$He and
$^{197}$Au (the latter with improved high-momentum tail of $n(k)$)
and compared with the experimental data taken from Fig. 6 of
\cite{Donn99}. The curves for $^{12}$C and $^{56}$Fe are not given
since they are similar, in agreement with the data.

\begin{figure}[th]
\includegraphics[width=10cm]{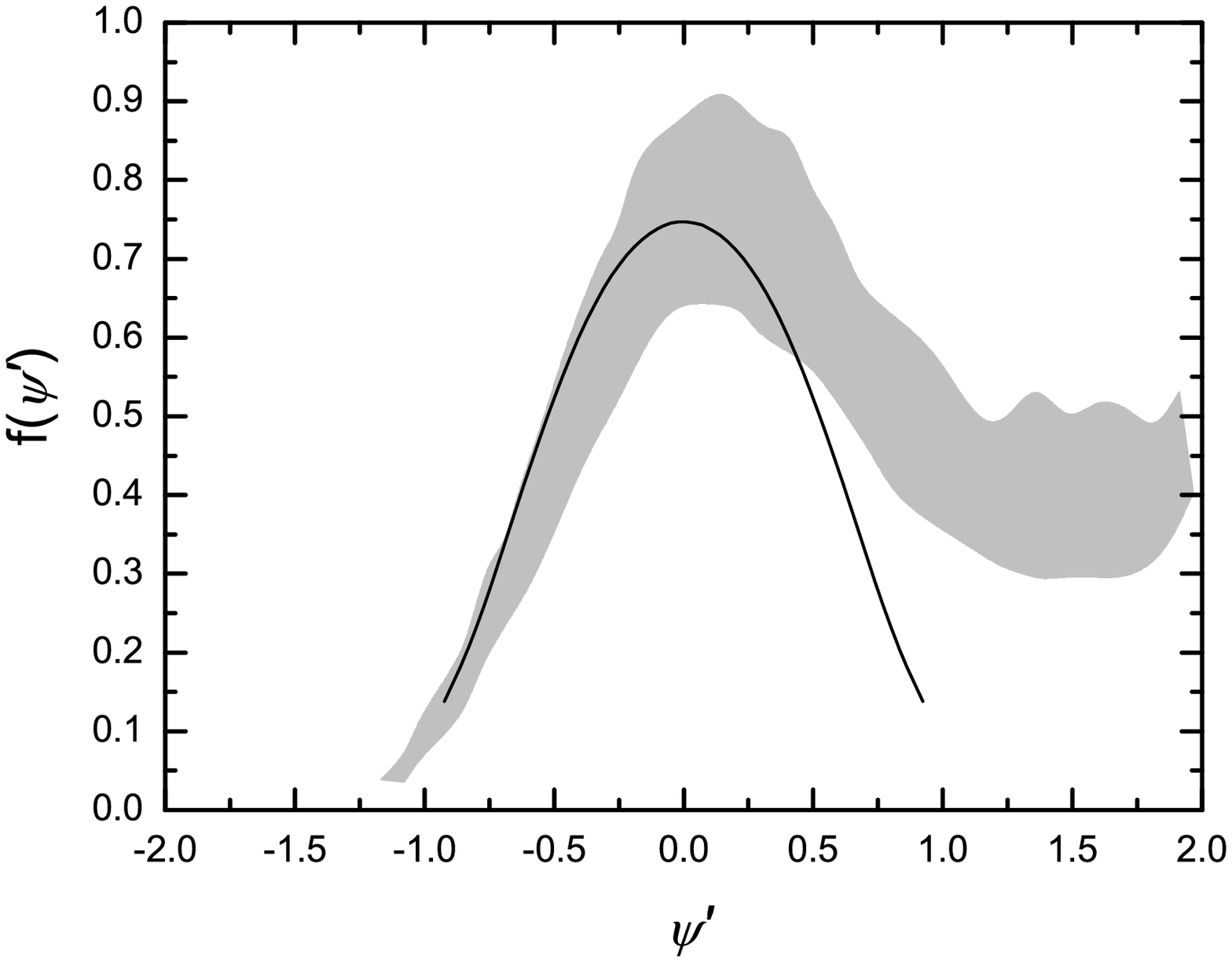}\\
\caption{Results of the CDFM for the superscaling function of
$^{12}$C at $q=500$ MeV/c (solid line) compared with the
experimental data (grey area) for $q$ in the interval from 500 to
600 MeV/c from \cite{Donn99}.}
\label{fig5}
\end{figure}

The result for the scaling function in the CDFM in the case of
$q$=500 MeV/c for $^{12}$C is given in Fig. \ref{fig5}. This
result is also in good agreement with the experimental data for
$q$ in the interval from 500 to 600 MeV/c given in Fig. 8 of
\cite{Donn99}.

One can see in Figs. \ref{fig2} and \ref{fig4} that the CDFM
results tend to overestimate the data in the interval $-1\leq
\psi^{\prime}\leq -0.5$. We note that the origin of this is
related to the predictions of the RFG model in the same region, as
can be seen in Fig. \ref{fig1}.

\section{Conclusions and final remarks}
The results of the present work can be summarized as follows:

i) We propose an extension of the RFG model to calculate the
scaling function $f(\psi^{\prime})$ in finite nuclei within the
Coherent Density Fluctuation Model (CDFM). In this model
$f(\psi^{\prime})$ is a weighted superposition of scaling
functions for relativistic Fermi gases with different densities.
The weight function is calculated using the known charge density
distributions in nuclei.

ii) We calculate the scaling function $f(\psi^{\prime})$ for
inclusive electron scattering for $^{4}$He, $^{12}$C, $^{27}$Al,
$^{56}$Fe and $^{197}$Au nuclei and for various values of the
transfer momentum $|{\bf q}|$=1650, 1000 and 500 MeV/c. The
results agree with the available experimental data at different
transferred momenta, and energies below the quasielastic peak
position, showing superscaling for negative values of
$\psi^{\prime}$ including also those smaller than -1. This is an
improvement over the RFG model predictions where the scaling
function becomes abruptly zero beyond $|\psi^{\prime}|=-1$.

iii) The sensitivity of the scaling function to the high-momentum
components of the momentum distribution is analyzed in detail,
especially so on the example of the $^{197}$Au nucleus.

iv) The scaling function obtained is symmetrical around
$\psi^{\prime}=0$ up to $|\psi^{\prime}|=(4ck_{F})^{-1}$. It would
also be interesting to search for models predicting an
assymetrical superscaling function $f(\psi^{\prime})$ as the
phenomenological one obtained by \cite{Jourdan96} and discussed in
Ref. \cite{Barb2003}.

It is shown in our work that the superscaling in nuclei can be
explained quantitatively on the basis of the similar behaviour of
the high-momentum components of $n(k)$ in light, medium and heavy
nuclei which is known to be due to the short-range and tensor
correlations in nuclei. This suggests an alternative path for
defining the weight function $F(x)$ within the generator
coordinate method: a path in which $F(x)$ is built up from a
phenomenological or a theoretical momentum distribution.

\acknowledgments One of the authors (A.N.A.) is grateful for warm
hospitality to the Faculty of Physics of the Complutense
University of Madrid and for support during his stay there to the
State Secretariat of Education and Universities of Spain
(Nº.Ref.SAB2001-0030). Four of the authors (A.N.A., M.K.G., D.N.K.
and M.V.I.) are thankful to the Bulgarian National Science
Foundation for partial support under the Contract No. $\Phi$-905.
This work was partly supported by funds provided by DGI of MCyT
(Spain) under Contracts BFM 2002-03562, BFM 2000-0600 and BFM
2003-04147-C02-01 and by the Agreement (2004 BG2004) between the
CSIC (Spain) and the Bulgarian Academy of Sciences.


\begin{thebibliography}{99}

\bibitem{Day90} D. Day, J.S. McCarthy, T.W. Donnelly, and I. Sick,
Ann. Rev. Nucl. Part. Sci. {\bf 40}, 357 (1990).

\bibitem{Sick80} I. Sick, D. Day, and J.S. McCarthy, Phys. Rev. Lett. {\bf 45},
871 (1980).

\bibitem{Pace82} E. Pace and G. Salme, Phys. Lett. {\bf B110}, 411 (1982).

\bibitem{Ciofi83} C. Ciofi degli Atti, E. Pace, and G. Salme, Phys. Lett. {\bf B127},
303 (1983); Phys. Rev. C {\bf 43}, 1155 (1991).

\bibitem{Ciofi99} C. Ciofi degli Atti and G.B. West, Phys. Lett.
{\bf B458}, 447 (1999).

\bibitem{West75} G. B. West, Phys. Rep. {\bf 18}, 263 (1975).

\bibitem{Donn991} T. W. Donnelly and I. Sick, Phys.Rev. Lett. {\bf 82}, 3212 (1999).

\bibitem{Donn99} T. W. Donnelly and I. Sick, Phys. Rev. C {\bf
60}, 065502 (1999).

\bibitem{Alber88} W.M. Alberico, A. Molinari, T. W. Donnelly, E.L. Kronenberg,
and  J.W. Van Orden, Phys. Rev. C {\bf 38}, 1801 (1988).

\bibitem{Barb98} M.B. Barbaro, R. Cenni, A. De Pace, T.W. Donnely,
and A. Molinari, Nucl. Phys. {\bf A643}, 137 (1998).

\bibitem{Mai2002} C. Maieron, T.W. Donnely, and I. Sick, Phys. Rev. C {\bf
65}, 025502 (2002).

\bibitem{Barb2003} M.B. Barbaro, J.A. Caballero, T.W. Donnely, and
C. Maieron, nucl-th/0311088 (2003).

\bibitem{Ant79} A.N. Antonov, V.A. Nikolaev, and I.Zh. Petkov,
Bulg. J. Phys. {\bf 6}, 151 (1979); Z.Phys. {\bf A297}, 257
(1980); {\it ibid.} {\bf A304}, 239 (1982).

\bibitem{Ant85} A.N. Antonov, V.A. Nikolaev, and I.Zh. Petkov,
Nuovo Cimento {\bf A86}, 23 (1985).

\bibitem{Ant88} A.N. Antonov, P.E. Hodgson, and I. Zh. Petkov, {\it Nucleon
Momentum and Density Distributions in Nuclei} (Clarendon Press,
Oxford, 1988).

\bibitem{Ant93} A.N. Antonov, P.E. Hodgson, and I. Zh. Petkov, {\it Nucleon
Correlations in Nuclei} (Springer-Verlag, Berlin-Heidelberg-New
York, 1993).

\bibitem{Ant89} A.N. Antonov, Chr.V. Christov, E.N. Nikolov, I.Zh. Petkov, and
P.E. Hodgson, Nuovo Cimento {\bf A102}, 1701 (1989); A.N. Antonov,
D.N. Kadrev, and P.E. Hodgson, Phys. Rev. C {\bf 50}, 164 (1994).

\bibitem{Grif57} J.J. Griffin and J.A. Wheeler, Phys. Rev. {\bf 108}, 311 (1957).

\bibitem{Wild77} K. Wildermuth and Y.C. Tang, {\it A Unified
Theory of the Nucleus} (Vieweg, Braunschweig, 1977).

\bibitem{Bauh80} W. Bauhoff, Ann. Phys. {\bf 130}, 307 (1980).

\bibitem{Dirac30} P.A.M. Dirac, Proc. Cambridge Phil. Soc. {\bf
26}, 376 (1930).

\bibitem{Burov98} V.V. Burov, D.N. Kadrev, V.K. Lukyanov, and Yu.S. Pol',
Phys. At. Nucl. {\bf 61}, 525 (1998).

\bibitem{Vries87} H. De Vries, C.W. De Jager, C. De Vries, At.Data Nucl.Data
Tables {\bf 36}, 495 (1987).

\bibitem{Patt2003} J.D. Patterson and R.J. Peterson, Nucl. Phys. {\bf A717}, 235 (2003).

\bibitem{Jourdan96} J. Jourdan, Nucl. Phys. {\bf A603}, 117 (1996).
\end{thebibliography}
\end{document}